\newcommand{\beq}{\begin{quote}}
\newcommand{\enq}{\end{quote}}
\newcommand{\be}{\begin{equation}}
\newcommand{\en}{\end{equation}}
\newcommand{\del}{\delta}

\newcommand{\grad}{\bigtriangledown}
\documentstyle[12pt]{article}
\begin{document}
\title{ Is Bohm's interpretation of quantum mechanics consistent?}
\date{ }
\author{}
\maketitle

\begin{abstract}
The supposed equivalence  of the conventional  interpretation of quantum mechanics with Bohm's interpretation  is generally
demonstrated  only in the coordinate representation.  It is shown, however,  that  in the momentum representation this equivalence
is not valid.

\end{abstract}
Recently, there has been a renewed interest in David Bohm's interpretation of {\it non-relativistic}  quantum mechanics (QM) \cite{bohm}-\cite{valentini}   and  many pedagogical papers on this topic have appeared\cite{matzkin}-\cite{goldstein}, while
on line,  arXiv.org  lists over 200 submissions on this topic  during the past ten years. 
   Bohm claimed that 
``as long as the present general form of Schr\"{o}dinger's equation is retained the physical results obtained with this  suggested
alternative  are precisely the same as those obtained with the  conventional  interpretation", and that 
his interpretation
``leads in all possible experiments to the same predictions as are obtained from the usual interpretation" \cite{bohm}.
Similar assertions  also have been made  in references \cite{matzkin}-\cite{goldstein}, but 
 this equivalence is   usually demonstrated  only in the coordinate representation,  while  the implications of Bohm's  interpretation in the momentum representation 
are usually  ignored.  While there have been some criticisms in the past of Bohm's  interpretation of QM, (for example,
see reference
\cite{keller},\cite{vankampen}), we give here an elementary proof
that the  momentum 
distribution in  this interpretation differs from that in  standard QM.  We show that  the definition of particle  velocity in this interpretation, 
 implies that the product of   mass times velocity  is not equal to momentum, which is  inconsistency
 with both classical and quantum mechanics.

In Bohm's interpretation of quantum mechanics,  the velocity  of a particle with mass $m$ is given by
\be
\label{vel}
\vec v_B=\vec \grad S/m
\en
where $S/\hbar $ is the phase of the wave function $\psi$
obtained by solving the time dependent  Schr\"{o}dinger equation.
According to Bohm, 
\be
\label{vel2}
\vec v_B = \frac{d \vec q}{dt},
 \en
 where $\vec q$ is the time dependent  coordinate for the position of the particle,  and Eq.\ref{vel} becomes 
 a first order differential equation
that determines $\vec q$ as a function of time $t$,  given its initial value \cite{second}. But it turns out that the product  $m \vec v_B $  is not equal to the canonical momentum $\vec p$,  because $\vec v_B$ does not correspond to the velocity $\vec v$,  that is determined
in quantum mechanics  by the operator
\be 
\vec v =-\frac{i \hbar}{m} \vec \grad_q =\frac{\vec p}{m}.
\en   
A proof of this relation is given in Appendix A. Setting
\be
\psi=R\; exp({iS/\hbar}),
\en
where $R$ is the amplitude of $\psi$, we obtain
\be
\vec v \: \psi= ( \vec \grad_q S/m -i\hbar \vec \grad_q R / m R)\psi.
\en
But  in  Bohm's definition of the particle velocity, Eq. \ref{vel},  only the first term on the right hand side of this equation appears.
The relevance of the second term can be illustrated  by  considering the mean values 
 $ <\vec v>$ and $< \vec v ^2>$ in this representation for $\psi$.  We have 
 \be
<\vec v >= \int d^3q\; \psi^{\dagger} \vec v \psi = \int d^3q  \vec R^2 \vec \grad S/m=<\vec v_B>, 
\en
and
\be
\label{second}
< \vec v^2>=\int d^3q\;  \psi^{\dagger} (\vec v)^2 \psi = < (\vec v_B)^2> + (\hbar/m)^2  \int d^3q\; (\vec \grad R)^2.
\en
Hence, Eq.\ref{second} implies that  the second moment of the velocity distribution in conventional quantum mechanics  differs from that obtained  in Bohm's interpretation of the particle velocity, Eq. \ref{vel}, by the appearance of the  additional term $(\hbar/m)^2<(\vec \grad R)^2 / R^2>$  on the right hand side of this equation.
Remarkably, this  discrepancy \cite{discrepancy} is not even mentioned in any  of the recent articles  on Bohm's interpretation of wave mechanics   \cite{matzkin}-\cite{bene}. 

To  get agreement with the mean value  $<\vec v^2>$ in quantum mechanics, Eq.\ref{second}, Bohm's  interpretation requires, in addition to the Bohmian particle velocity $\vec v_B$ given by Eq.\ref{vel}, 
the  existence  of  an {\it ad hoc}  random velocity  
\be
\vec v_o=\frac{\hbar}{mR} \vec \grad R,
\en
with vanishing mean value, Originally, such a contribution was introduced with an undetermined coefficient 
  as a {\it random} velocity by D. Bohm and J. P. Vigier \cite{mismatch}, who named 
 it  an ``osmotic velocity", after a term introduced by Einstein to describe the chaotic  Brownian motion. But  now such a term has been abandoned  in  discussions of Bohmian mechanics.
 
 In particular, for stationary solutions of the Schr\"{o}dinger, the phase $S=0$, and Bohm's interpretation leads to the conclusion  that
 the particle velocity  vanishes in such a state. This conclusion is explained  by invoking 
 a {\it quantum force}  due to a  non-local {\it quantum potential} that supposedly  balances the force due to the conventional  potential that gives
 rise to the stationary solution. This non-classical force  appears   
 when the acceleration $ d^2 \vec q/ dt^2$  is calculated by taking the time derivative of  Eqs. \ref{vel} and \ref{vel2}. But this result  contradicts the fact that in quantum mechanics
  the  velocity or momentum distribution for stationary solutions,  given by the absolute square of  the Fourier transform of $\psi$ in coordinate space,  is not a delta function  at  $\vec v =0$,  as is implied by Bohm's interpretation.
  
  The  trajectories   obtained  by integrating  Bohm' s first order differential  equation for the particle coordinate $\vec q$, Eq. \ref{vel2}, 
  correspond to   {\it pathlines} associated with the probability distribution $\rho=|\psi|^2$  which satisfies, like a normal fluid of density
  $\rho$, the continuity equation,
  \be
  \frac{\partial \rho}{\partial t} + \vec \grad_q .\vec j =0
  \en
  where $\vec j =\vec v_B \rho$ is the associated current.
   While pathlines provide a visualization of a fluid flow,  these lines do not correspond to  the actual motion of the particles composing the fluid
   that  also can  have a random component.
  Likewise, Bohmian  pathlines   serve to visualize the evolution of the probability distribution in quantum mechanics,   
  but  do not correspond to actual trajectories of elementary particles.
  
  Recently,  experiments have been made with  water droplets surfing on the waves produced by the Faraday instability  on the surface of an oscillating tank filled
  with a fluid \cite{couder}.  The motion of these droplets mimics  the suggestion of de Broigle  and of Bohm  that elementary particles are likewise ``piloted"  
  by the $\psi$ function of wave mechanics. In particular, it is claimed that when the waves propagate  through two slits, or are
  confined in a ``corral",  the droplets satisfy  statistics that are similar to those observed for particles in  quantum mechanics \cite{bohr}. But  such experiments only 
  demonstrate   the  universality of wave propagation,  and the associated pathlines,   whether governed by the equations of fluid mechanics,  quantum mechanics,  or of  other sources of
  waves in physics.
    
  \subsection* {Appendix A. The relation between velocity and momentum  in non-relativist quantum mechanics}
  
In  quantum mechanics,   the velocity  $ \vec  v$,   like the  position $\vec q$  and the momentum $\vec p$,   is an operator. It is 
{\it defined} by the relation
\be
\label{velocity}
 \vec v= \frac{i}{\hbar} [ H,\vec q \;],
\en
where $H$ is the hamiltonian operator, and $[a,b]=ab-ba$ is the commutator of the operators 
$a$ and $b$.  In non-relativistic quantum mechanics,
\be
\label{hamiltonian}
H=-\frac{\hbar^2}{2m}\grad_{ q}^2 + V(\vec q),
\en
corresponding to the time dependent Schr\"{o}dinger equation
\be
\label{sch}
i\hbar \frac{\partial \psi}{dt}= H\psi
\en
Hence, substituting  this expression for $H$ in Eq.\ref{velocity}, one finds that the velocity operator is given by
\be
\label{vel3}
\vec v=\frac{\vec p}{m}
\en
where 
\be
\vec p=-i\hbar \vec \grad_q
\en
is the momentum operator.

For an alternative derivation of the connection between  the velocity and momentum operators, Eq.\ref{vel3}, 
 that does not presuppose the 
Schr\"{o}dinger equation, Eqs.\ref{hamiltonian} and \ref{sch}, consider the commutation relation Eq.\ref{velocity} 
 for the Hamiltonian of a free particle  $H_o={\vec p}^2 /2m$.
Then, according to the definition of velocity, Eq.\ref{velocity}, 
\be
 v_i=\frac{i}{2\hbar m} (p_j [p_j,q_i]+[p_j,q_i]p_j), 
 \en
 and  substituting  the Heisenberg-Born commutation relation
 \be
 [p_j,q_i]=-i\hbar \del_{i,j} 
 \en
 leads again  to  Eq. \ref{vel3}.

\end{document}